\documentclass[11pt,3p,review,authoryear]{elsarticle}

\journal{International Journal of Forecasting}
\biboptions{longnamesfirst}
\usepackage[many]{tcolorbox}
\usepackage{colortbl}
\usepackage{ulem} 
\definecolor{codegreen}{rgb}{0,0.6,0}

\usepackage[colorlinks=true, linkcolor=blue, citecolor=blue, urlcolor=blue]{hyperref}
\usepackage{amsthm}
\usepackage{booktabs}
\usepackage{tikz}
\usepackage{algorithm, algpseudocode}
\usetikzlibrary{positioning}
\usetikzlibrary{arrows.meta}
\usepackage{multirow}
\usepackage[T1]{fontenc}
\usepackage{lmodern}
\usetikzlibrary{arrows.meta,calc}
\usepackage{amsfonts}
\usepackage[mathscr]{euscript}
\usepackage{array}
\usepackage{subcaption}
\usepackage{soul}

\newtheorem{definition}{Definition}

\newcommand{\Bbf}{\mathbf{B}}
\newcommand{\Ubf}{\mathbf{U}}
\newcommand{\Ybf}{\mathbf{Y}}
\newcommand{\Sbf}{\mathbf{S}}
\newcommand{\Abf}{\mathbf{A}}
\newcommand{\Wbf}{\mathbf{W}}
\newcommand{\Zbf}{\mathbf{Z}}
\newcommand{\Kbf}{\mathbf{K}}

\newcommand{\xbf}{\mathbf{x}}
\newcommand{\bbf}{\mathbf{b}}
\newcommand{\ubf}{\mathbf{u}}
\newcommand{\ybf}{\mathbf{y}}
\newcommand{\zbf}{\mathbf{z}}

\newcommand{\rr}{\mathbb{R}}
\newcommand{\mm}{\mathcal{M}}

\newcommand{\yH}{\widehat{\mathbf{y}}}
\newcommand{\bH}{\widehat{\mathbf{b}}}
\newcommand{\uH}{\widehat{\mathbf{u}}}
\newcommand{\YH}{\widehat{\mathbf{Y}}}
\newcommand{\UH}{\widehat{\mathbf{U}}}
\newcommand{\BH}{\widehat{\mathbf{B}}}
\newcommand{\Sigmabf}{\mathbf{\Sigma}}
\newcommand{\SigmaT}{\widetilde{\mathbf{\Sigma}}}
\newcommand{\SigmaH}{\widehat{\mathbf{\Sigma}}}

\newcommand{\yT}{\widetilde{\mathbf{y}}}

\newcommand{\bT}{\widetilde{\mathbf{b}}}

\newcommand{\BT}{\widetilde{\mathbf{B}}}

\newcommand{\nutil}{\widetilde{\nu}}
\newcommand{\nuhat}{\widehat{\nu}}
\newcommand{\pitil}{\widetilde{\pi}}
\newcommand{\pihat}{\widehat{\pi}}

\DeclareMathOperator*{\argmin}{arg\,min}

\newcommand{\figref}[1]{Fig.~\ref{#1}}
\newcommand{\tabref}[1]{Tab.~\ref{#1}}
\newcommand{\eqtref}[1]{Eq.~\eqref{#1}}

\begin{document}

\begin{frontmatter}

\title{Nonlinear Probabilistic Forecast Reconciliation }

\author[isaac,idsia]{Anubhab Biswas\corref{cor}}
\cortext[cor]{Corresponding author}
\ead{anubhab.biswas@supsi.ch}
\author[idsia]{Lorenzo Zambon}
\author[isaac]{Lorenzo Nespoli}
\author[idsia]{Giorgio Corani}
\address[isaac]{ISAAC, SUPSI, Institute for Applied Sustainability in the Built Environment, CH-6850, Mendrisio, Switzerland}
\address[idsia]{IDSIA (USI-SUPSI), Dalle Molle Institute for Artifical Intelligence, CH-6892, Lugano, Switzerland}

\begin{abstract}
    Forecast reconciliation adjusts independently generated forecasts across multiple time series so that they satisfy some known constraints. While probabilistic forecast reconciliation is well established for linear constraints, some practical forecasting problems involve nonlinear relationships among time series. In this paper, we address probabilistic  forecast reconciliation with nonlinear constraints for the first time. We extend both reconciliation via projection and conditioning to the case of nonlinear constraints. The former approach reconciles forecast samples by mapping them onto the nonlinear coherent manifold. The latter approach conditions the joint forecast distribution on the constraints; the reconciled distribution is approximated with an algorithm based on the Unscented Kalman Filter (UKF). We evaluate both methods on synthetic and real datasets. Empirically, both reconciliation approaches generally improve forecast accuracy. The UKF-based approach achieves the best overall performance, is consistently retained in the Model Confidence Set — either alone or alongside other reconciliation methods — and is substantially faster than the projection-based approach.
\end{abstract}


\end{frontmatter}

\section{Introduction}

Time series are often related by linear constraints of aggregation. For example, the total sales of a store are equal to the sum of the sales of all items in that store. This relationship should  be preserved when forecasting: the forecast of the total  sales should equal the sum of the forecasts for the individual items.
The forecasts are called \textit{coherent} when they satisfy these constraints, and \textit{incoherent} when they do not. The \textit{base forecasts}, which are created independently for each time series, are generally incoherent. \textit{Forecast reconciliation} adjusts the base forecasts making them coherent.

The first methods for forecast reconciliation can be interpreted as  projecting the base point forecast onto the linear coherent subspace \citep{ATHANASOPOULOS2009146, HYNDMAN20112579}, which is the subspace where the constraints are satisfied. In particular, 
the minT algorithm \citep{Wickramasuriya03042019} yields the projection that minimizes the expected squared error of the coherent forecast.

Such methods however reconcile only the point forecasts.
The most advanced approaches instead perform probabilistic reconciliation, i.e., they provide  a probability density (or a mass function) on the linear coherent subspace. Earlier studies on probabilistic reconciliation include \citet{Taieb02012021}, where an empirical copula based approach for reconciliation which uses permutation of the samples was studied. Later  two main approaches emerged. The first extends \textit{reconciliation via projection} to the probabilistic setting \citep{PANAGIOTELIS2023693, Wickramasuriya02012024, girolimetto2024point}. The second is \textit{reconciliation via conditioning}, where the incoherent distribution of the base forecasts is sliced on the coherent subspace. This approach  has been applied to continuous, discrete, and mixed-type time series \citep{zambon2024efficient, ZAMBON20241438, pmlr-v244-zambon24a}.

So far, most  literature  has focused on linear constraints, although nonlinear constraints arise  in many forecasting problems.  Socio-demographic rates provide an important example: jointly forecasting the population, the number of employed persons, and the employment rate results in time series linked by nonlinear constraints \citep{girolimetto2025forecastreconciliationnonlinearconstraints}. Other examples include power system variables constrained by power flow equations \citep{MehtaMolzahnTuritsyn2016}, water network variables linked by tank dynamic constraints \citep{SinghKekatos2020}, and chemical process variables related through reactor mass and energy balance constraints \citep{RawlingsEkerdt2012,Bequette2023}. In these cases the forecasts  should generally satisfy both linear and nonlinear constraints. Only recently, reconciliation via projection has been extended to nonlinear constraints \citep{nespoli2026nonlinear, girolimetto2025forecastreconciliationnonlinearconstraints}; however, these approaches are limited to the reconciliation of point forecasts. 

In this paper, we address, for the first time, probabilistic forecast reconciliation under nonlinear constraints, meaning that at least one of the constraints is nonlinear. In this setting, the reconciled coherent distribution lies on a nonlinear manifold rather than on a linear subspace. We extend both reconciliation via \textit{projection} and reconciliation via \textit{conditioning} to this setting. 
The proposed projection-based approach draws samples from the incoherent base forecast distribution and maps each sample onto the coherent manifold. Its main drawback is that reconciling each sample requires solving a numerical optimization problem, resulting in high computational costs. 
For reconciliation via conditioning we instead approximate the reconciled distribution with an algorithm based on the Unscented Kalman Filter \citep{JulierUhlmann2004}. In our experiments, this approach yields slightly better forecasts than projection, with substantially lower computation time.

The paper is organized as follows. Section 2 introduces the notations.
Section 3 presents probabilistic reconciliation under nonlinear constraints, both via projection and via conditioning. Section 4 evaluates the proposed methods on synthetic data and real case studies involving socio-demographic ratios of Switzerland and tourism ratios of Australia. Section 5 concludes and highlights the directions for future research.

\section{Notation}

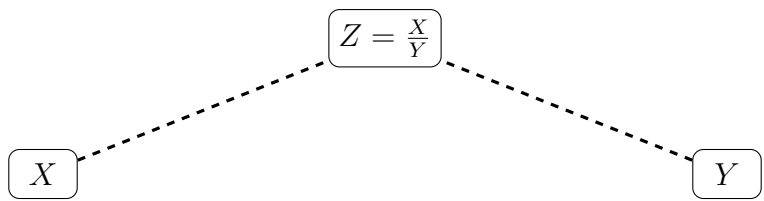
\begin{figure}[h!]
\centering
\begin{tikzpicture}[scale=0.9, every node/.style={rectangle,draw,align=center,rounded corners=4,font=\large, minimum height=0.65cm, minimum width=0.9cm},
]
\node (R) at (0,5) {$Z = \frac{X}{Y}$};
\node (P) at (-5,3) {$X$};
\node (I) at (5,3) {$Y$};

\draw[dashed, very thick] (P) -- (R);
\draw[dashed, very thick] (I) -- (R);
\end{tikzpicture}
   \caption{Example of time series with nonlinear constraints; the free time series are $X$ and $Y$, the constrained time series $Z$ is the ratio of the two. 
   \label{fig:nlc}}
\end{figure}

Let us consider the time series $\Bbf_t = [B^1_t, \dots, B^{n_b}_t]^T \in \rr^{n_b}$  and $\Ubf_t = [U^1_t, \dots, U^{n_u}_t]^T \in \rr^{n_u}$, such that, at each time $t=1,\dots,T$: 
\begin{equation}
    \Ubf_t = f_u(\Bbf_t),
\end{equation}
for some function $f_u : \rr^{n_b} \to \rr^{n_u}$.
This formulation generalizes the classic framework of hierarchical time series, in which $f_u(\Bbf_t) = \Abf \Bbf_t$, and $\Abf \in \rr^{n_u \times n_b}$ is the aggregation matrix.
We refer to $\Bbf_t$ and $\Ubf_t$ respectively as \textit{free} and \textit{constrained} time series, and we call $f_u$ the \textit{free-to-constrained} (FTC) function.
We show an example in \figref{fig:nlc}, where $f_u(X_t,Y_t) = X_t \,/\, Y_t$.
Note that the choice for the set of free time series is not unique; in this case, we could equivalently set $X_t$ as the constrained series and $f_u(Y_t, Z_t) = Y_t \cdot Z_t$.
We then define the full vector of time series as
\begin{equation}\label{eq:full_set_coh}
    \Ybf_t =
    \begin{bmatrix}
        \Ubf_t \\
        \Bbf_t
    \end{bmatrix}
    =
    \begin{bmatrix}
        f_u(\Bbf_t) \\
        \Bbf_t
    \end{bmatrix} =   f(\Bbf_t),
\end{equation}
where $f : \rr^{n_b} \to \rr^n$ is the \textit{free-to-all} (FTA) function. At each time $t$, the vector $\Ybf_t$ lies on the nonlinear manifold
\begin{equation}\label{eq:manifold}
    \mathcal{M} := \{ f(\bbf) : \bbf \in {\mathscr{B} \subseteq} \mathbb{R}^{n_b} \},
\end{equation}
where $\mathscr{B}\subseteq \mathbb{R}^{n_b}$ denotes the admissible domain of the free time series; in the example of \figref{fig:nlc}, $\mathscr{B} = \{(x,y) : x\in \rr,\; y \in \rr \setminus \{0\}\} \subseteq \rr^2$. The manifold $\mm$ and the FTA $f(.)$ are respectively the generalizations of the linear coherent subspace $\mathcal{S}$ and the summing matrix $\Sbf$ used in linear reconciliation. We summarize the notation in \tabref{tab:notations}, where we also compare it with its linear counterpart; we keep it identical when possible. 

\begin{figure}
    \centering
    \includegraphics[width=0.45\linewidth]{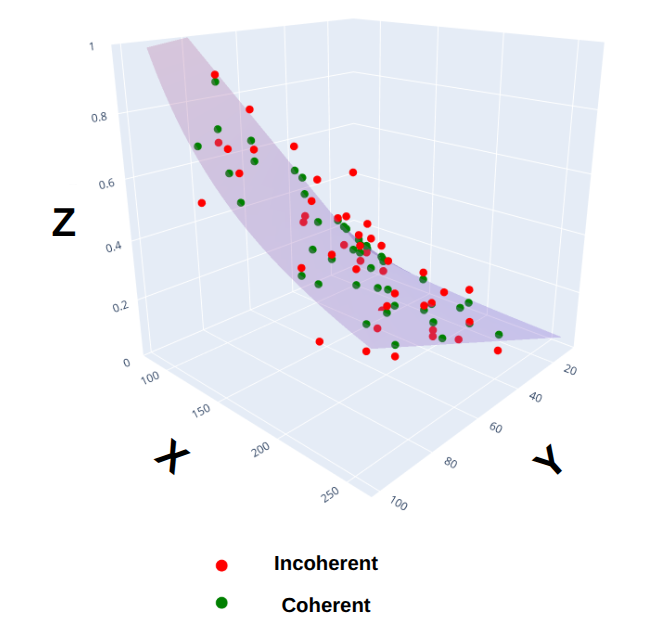}
    \caption{Sample-based representation of the coherent (green) and incoherent (red) probability distributions for the ratio constraint of \figref{fig:nlc}. 
    The purple surface represents the coherent manifold.}
    \label{fig:PNLR}
\end{figure}


\begin{table}[h!]
    \centering
    \begin{tabular}{ll@{\hspace{1.5cm}}ll}
     \toprule
     Linear Reconciliation Object & Notation & Generalization & Notation \\
     \midrule
     Bottom time series & $\Bbf_t$ & Free time series & $\Bbf_t$ \\
     Upper time series & $\Ubf_t$ & Constrained time series & $\Ubf_t$\\
     Coherent linear subspace & $\mathcal{S}$ & Coherent manifold & $\mathcal{M}$ \\
     Aggregation matrix & $\Abf$ & Free-to-constrained (FTC) function & $f_u(\cdot)$ \\
     Summing matrix & $\Sbf$ & Free-to-all (FTA) function & $f(\cdot)$ \\
     No. of bottom series & $n_b$ & No. of free series & $n_b$ \\
     No. of upper series & $n_u$ & No. of constrained series & $n_u$ \\
     No. of series & $n$ & No. of series & $n$ \\
     \bottomrule
    \end{tabular}
     \caption{A generalized notation of forecast reconciliation
     }
    \label{tab:notations}
\end{table}



Following \citet{PANAGIOTELIS2023693}, we denote the forecast distribution of the free time series $\Bbf_t$ by the probability triple $(\mathbb{R}^{n_b},\mathcal{F}_{\rr^{n_b}},\nu_{n_b})$, where $\mathcal{F}_{\rr^{n_b}}$ denotes the Borel $\sigma$-algebra on $\rr^{n_b}$. 
We extend the definition of coherence of probabilistic forecasts given by \citet{PANAGIOTELIS2023693} to the case of nonlinear constraints.
%
\begin{definition}[\textbf{Coherent Probability Triple}]\label{def:coherence}
    A probability triple $(\mathcal{M},\mathcal{F}_{\mathcal{M}},\breve{\nu}_{\mathcal{M}})$ is coherent with the free-level probability triple $(\mathbb{R}^{n_b},\mathcal{F}_{\mathbb{R}^{n_b}},\nu_{n_b})$ if
    \begin{equation}\label{eq:def_coherence}
        \breve{\nu}_{\mathcal{M}}(f(\mathcal{B})) = \nu_{n_b}(\mathcal{B}) \quad \forall \mathcal{B} \in \mathcal{F}_{\mathbb{R}^{n_b}}.
    \end{equation}
\end{definition}
As noted by \citet{zambon2024efficient}, Eq.~\eqref{eq:def_coherence} is equivalent to saying that $\breve{\nu}_{\mathcal{M}}$ is the push-forward of $\nu_{n_b}$ via $f$.
In \figref{fig:PNLR}, we show a sample-based representation of coherent and incoherent densities for the example of \figref{fig:nlc}.
For simplicity, in the following we drop the triple notation and denote the joint forecast distribution for all the time series by $\nuhat$; moreover, we assume it to be continuous and denote its density on $\rr^n$ by $\pihat$.


\section{Probabilistic Methods for Nonlinear Reconciliation}

We now show how to extend reconciliation via projection \citep{PANAGIOTELIS2023693} and reconciliation via conditioning \citep{zambon2024efficient, Corani2021} to the case of nonlinear constraints.

\subsection{Reconciliation via projection}
\label{sec:rec_via_proj}

In the linear case, \citet{PANAGIOTELIS2023693} defines probabilistic forecast reconciliation through the push-forward of measures. Given a base forecast distribution $\nuhat$ on the space $\rr^n$, the reconciled measure $\nutil$ is defined as the push-forward $\nutil = \psi_{\#} \nuhat$ under a reconciliation map $\psi: \rr^n \to \mathcal{S}$, where $\mathcal{S}$ is the coherent subspace: for any measurable set $\mathcal{B} \subseteq \mathcal{S}$, the reconciled probability is given by $\nutil(\mathcal{B}) = \nuhat(\psi^{-1}(\mathcal{B}))$.

In the linear literature, $\psi$ is commonly defined as an orthogonal projection onto $\mathcal{S}$ with respect to some distance on $\rr^n$.
Given a positive definite matrix $\Wbf \in \rr^{n \times n}$, the distance is derived from the induced norm $\|\xbf\|_{\mathbf{W}} = \sqrt{\xbf'\,\Wbf^{-1}\, \xbf}$.
Then, $\psi$ maps each point in $\rr^n$ to its nearest point on $\mathcal{S}$:
$\psi(\yH) = \argmin_{\xbf \in \mathcal{S}} \,\, \|\yH - \xbf\|_{\Wbf}$.
When $\Wbf$ is an estimate of the covariance matrix of the base forecast errors, this  corresponds to the Minimum Trace (MinT) projection \citep{Wickramasuriya03042019}.
\cite{Wickramasuriya03042019} and \cite{Wickramasuriya02012024}  discuss
the optimality properties of MinT for point forecasts and, under the Gaussian assumption, for probabilistic forecasts.

\citet{nespoli2026nonlinear} and \citet{girolimetto2025forecastreconciliationnonlinearconstraints} reconcile point forecasts subject to nonlinear constraints by projecting them to their nearest point on the coherent manifold $\mm$. Since $\mm$ is nonlinear, this nearest-point mapping is nonlinear. We extend this idea to probabilistic reconciliation by defining the reconciled distribution as the push-forward of the base forecast distribution through the nonlinear nearest-point map.
However, for a nonlinear manifold $\mm$, the nearest-point map may not exist or may not be unique \citep{nespoli2026nonlinear}. 
We thus assume that $\mm$ is non-empty and closed and that $\nuhat$ is continuous; under these assumptions, $\psi$ is measurable and uniquely defined for $\nuhat$-almost every $\yH$, making the following definition well-posed. A more detailed discussion is provided in \ref{app:wellposed}.

\begin{definition}[\textbf{Reconciled distribution via projection}]
Given a norm $\|\cdot\|_{\Wbf}$ on $\rr^n$ induced by a positive definite matrix $\Wbf\in\rr^{n \times n}$, 
let $\psi: \rr^n \to \mm$ be the nearest-point projection map:
\begin{equation} \label{eq:nearest_point_proj_nonlinear}
\psi(\yH) = \argmin_{\xbf \in \mm} \,\, \|\yH - \xbf\|_{\Wbf}.
\end{equation}
The reconciled distribution via projection is defined as $\nutil = \psi_{\#} \nuhat$, the push-forward of the base forecast distribution $\nuhat$ via $\psi$, specified by
\begin{equation}
\nutil(\mathcal{B}) = \nuhat(\psi^{-1}(\mathcal{B})), \quad \text{for any measurable } \mathcal{B} \subseteq \mm.    
\end{equation}
\end{definition}
As in the linear case \citep{PANAGIOTELIS2023693}, we obtain samples from the reconciled distribution $\nutil$ by mapping samples drawn from the base distribution $\nuhat$ via $\psi$:
\begin{equation*}
    \yT^{(i)} = \psi\left(\yH^{(i)}\right), \quad \text{for all } i=1,\dots,M
\end{equation*}
where $\left(\yH^{(i)}\right)_{i=1\dots M} \sim \nuhat$.
Since $\psi$ is uniquely defined for $\nuhat$-almost every point, each sample $\yH^{(i)}$ lies in
its domain with probability one, so this procedure is almost surely well defined.
However, $\psi$ in general does not admit a closed-form expression.
Hence, for each sample $\yH^{(i)}$ we numerically solve Eq.~\eqref{eq:nearest_point_proj_nonlinear} via  optimization,
using the Newton-Raphson algorithm as in \citet{nespoli2026nonlinear}.  
A drawback of this approach is thus its computational cost, which increases linearly with the number of samples $M$. 
We consider three specifications for the matrix $\Wbf$, each defining a different projection:
\begin{enumerate}
    \item \textbf{OLS}: we set $\Wbf$ equal to the identity matrix; this corresponds to the Euclidean distance in $\rr^n$.
    This specification does not account for differences in scale across time series.
    \item \textbf{WLS}: we set $\Wbf$ as a diagonal matrix, whose entries are the estimated variances of the base forecast errors.
    This specification accounts for differences in scale across the time series.
    \item \textbf{Full}: we set $\Wbf$ as the full covariance matrix of the base forecast errors, estimated via shrinkage as in \citet{Wickramasuriya03042019}.
    This specification accounts for both differences in base forecast error variances and correlations among the base forecast errors.
\end{enumerate}

\subsection{Reconciliation via conditioning}

This approach conditions the joint base forecast distribution on the  constraints \citep{zambon2024efficient}.
In the following, we assume that $\nuhat$ is a continuous distribution, and we denote its density by  $\pihat$.

\begin{figure}[h!]
    \centering
    \begin{subfigure}{.5\textwidth}
  \includegraphics[height=5cm]{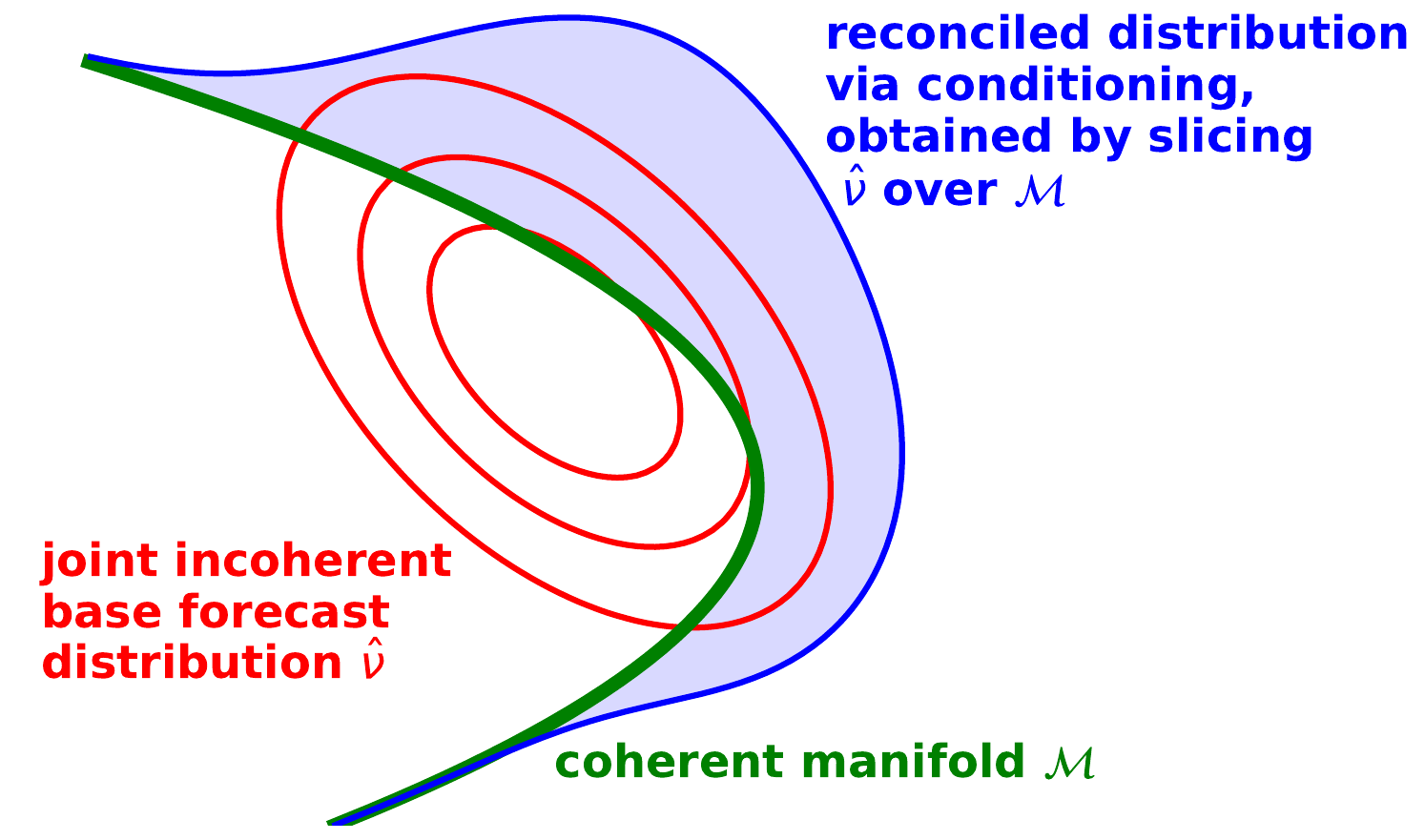}
\end{subfigure}%
\begin{subfigure}{.5\textwidth}
  \hspace{2mm}
  \includegraphics[height=6cm]{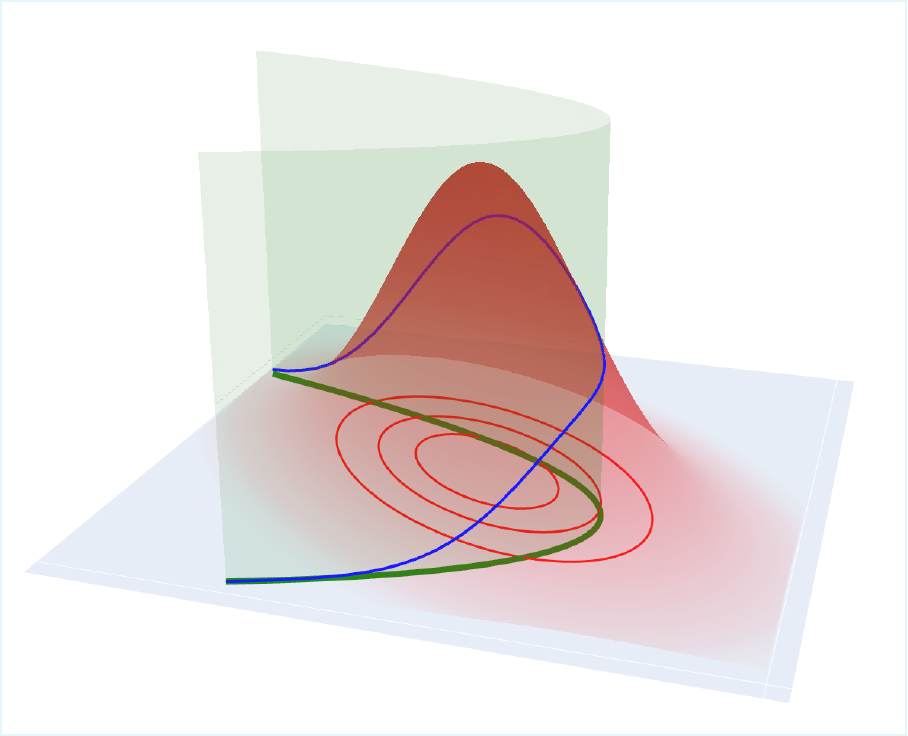}
\end{subfigure}
\caption{Representation of nonlinear probabilistic reconciliation via conditioning
}
\label{fig:rec_cond_3dplot}
\end{figure}

\begin{definition}[\textbf{Reconciled distribution via conditioning}]
The reconciled distribution via conditioning of the free time series is defined as the probability distribution $\nutil_B$ on $\rr^{n_b}$ with density 
\begin{equation}\label{eq:rec_distr_cond_bottom}
\pitil_B(\bbf) \propto \pihat(f_u(\bbf),\,\bbf).     
\end{equation}
The reconciled distribution via conditioning $\nutil$ for the entire set of time series is then defined as the push-forward of $\nutil_B$ via the FTA map:
\begin{equation}\label{eq:rec_distr_cond_all}
\nutil = f_{\#} \nutil_B.
\end{equation}
\end{definition}
We now intuitively explanation of the definition, leaving the details in \ref{app:derivation_rec_cond}. 
Let ${\YH = \left[\UH^T,\, \BH^T\right]^T}$ be a random vector distributed as the base forecast distribution $\nuhat$.
Eq.~\eqref{eq:rec_distr_cond_bottom} provides the density of the conditional distribution of $\BH$, given the constraints:
\begin{equation}
\BH\;\Big| \left(\UH - f_u(\BH) = 0\right) \;\;\sim\;\; \nutil_B.
\end{equation}
Then, Eq.~\eqref{eq:rec_distr_cond_all} follows  from
the definition of coherent forecast distribution (Def.~\ref{def:coherence}).
Analogously to the linear case, reconciliation via conditioning can be interpreted as slicing the incoherent joint density over the coherent manifold $\mm$ (Fig.~\ref{fig:rec_cond_3dplot}).

\paragraph{Approximating the reconciled distribution via the UKF}
Since the density in Eq.~\eqref{eq:rec_distr_cond_bottom} is  not available in closed-form, we  resort to an approximation based on the Unscented Kalman Filter \citep[UKF, ][]{JulierUhlmann2004}. A modern  tutorial on the UKF is given by \citet[Chap.~10]{Labbe2020Kalman}.  

We assume that the base forecasts of the free and
constrained series are independent, so that
Eq.~\eqref{eq:rec_distr_cond_bottom} factorizes as
\begin{equation}\label{eq:rec_distr_cond_product}
\pitil_b(\bbf)\ \propto\  \pihat_b(\bbf) \; \pihat_u\big(f_u(\bbf)\big),   
\end{equation}
where $\pihat_b$ and $\pihat_u$ are the densities of the distributions of $\BH$ and $\UH$, respectively.
We further assume that these densities are multivariate Gaussian, 
and we estimate their covariance matrices via shrinkage from the in-sample residuals; this is standard in probabilistic forecast reconciliation \citep{Corani2021, PANAGIOTELIS2023693, Wickramasuriya02012024}.
Since $f_u$ is nonlinear, however, the product in Eq.~\eqref{eq:rec_distr_cond_product} does not yield a Gaussian density.
We proceed in three steps (Algorithm~\ref{alg:ukf}); the computations are detailed in \textit{Appendix~S1} of the supplementary material.

\begin{enumerate}[(a)]
    \item \textit{Unscented transform}. 
    We represent $\pihat_b = \mathcal N(\bH,\,\SigmaH_b)$ by a set of $2n_b+1$ weighted \textit{sigma points}, 
    deterministically defined as in \citet{JulierUhlmann2004} so that their mean and covariance equal $\bH$ and $\SigmaH_b$.
    Then, we propagate the sigma points through the FTC map $f_u$ and compute, by moment matching, a Gaussian approximation of the joint distribution of $\BH$ and $f_u(\BH)$.
    \item \textit{Conditioning}.
    The distribution $\pitil_b$ in Eq.~\eqref{eq:rec_distr_cond_product} can be interpreted as the conditional distribution of $\BH$, given a noisy observation of $f_u(\BH)$, where $\uH$ is the observed value and $\SigmaH_u$ the observation noise.
    Since we approximate the joint distribution of $\BH$ and $f_u(\BH)$ with a Gaussian, this conditioning can be done in closed form,  yielding the Gaussian approximation $\mathcal N(\bT,\SigmaT)$ of the reconciled distribution $\pitil_b$. This conditioning corresponds to the  update step \citep[Chap.~10.5.2]{Labbe2020Kalman} of the Unscented Kalman filter.   
    In the case of linear constraints, the relation between the state-update step of the Kalman filter and probabilistic forecast reconciliation is discussed by \citet{Corani2021}.
    \item \textit{Sampling}. 
    From Eq.~\eqref{eq:rec_distr_cond_all}, we obtain coherent samples for all series by drawing samples from $\mathcal N(\bT,\SigmaT)$ and mapping them onto $\mm$ via the FTA map $f$.
\end{enumerate}

\begin{algorithm}[h!]
\caption{Approximating the reconciled distribution via the UKF}
\label{alg:ukf}
\begin{algorithmic}[1]
\Require mean and covariance of the free and constrained base forecast
         distributions: $\bH,\,\SigmaH_b,\,\uH,\,\SigmaH_u$
\Ensure  coherent samples $\{\yT^{(j)}\}_{j=1}^{M}$
\Statex (a) Unscented transform
\State generate sigma points $\bbf^{(i)}$ and weights
       $w_m^{(i)},w_c^{(i)}$ from $\mathcal N(\bH,\SigmaH_b)$
       \Comment{sigma points of $\pihat_b$}
\State $\ubf^{(i)} \gets f_u\!\left(\bbf^{(i)}\right)$ \;\;for all $i$
       \Comment{propagate sigma points}
\State $\bar\ubf \gets \sum_i w_m^{(i)}\,\ubf^{(i)}$
\Comment{approx. mean of $f_u(\BH)$}
\State $\Sigmabf_{uu} \gets \sum_i w_c^{(i)}
       \!\left(\ubf^{(i)}-\bar\ubf\right)\!\left(\ubf^{(i)}-\bar\ubf\right)^{\!\top}$
       \Comment{approx. cov.\ of $f_u(\BH)$}
\State $\Sigmabf_{bu} \gets \sum_i w_c^{(i)}
       \!\left(\bbf^{(i)}-\bH\right)\!\left(\ubf^{(i)}-\bar\ubf\right)^{\!\top}$
       \Comment{approx. cross-cov.\ of $\BH$ and $f_u(\BH)$}
\Statex (b) Conditioning
\State $\Kbf \gets \Sigmabf_{bu}\,\big(\Sigmabf_{uu} + \SigmaH_u\big)^{-1}$ \Comment{Kalman gain}
\State $\bT \gets \bH + \Kbf\!\left(\uH-\bar\ubf\right)$
       \Comment{reconciled free mean}
\State $\SigmaT_b \gets \SigmaH_b - \Kbf\,\big(\Sigmabf_{uu} + \SigmaH_u\big)\,\Kbf^{\top}$
       \Comment{reconciled free covariance}
\Statex (c) Sampling
\State draw $\bT^{(j)}\sim\mathcal N(\bT,\,\SigmaT_b)$
       for all $j=1,\dots,M$
       \Comment{reconciled free samples}
\State $\yT^{(j)} \gets f\!\left(\bT^{(j)}\right)$ \;\;for all $j$
       \Comment{map samples onto $\mm$}
\end{algorithmic}
\end{algorithm}

\subsection{Limitations}
The proposed methods assume the base forecast distribution $\nuhat$ to be continuous.
However, some applications may involve discrete or mixed predictive distributions. For instance, consider forecasting two low-count free time series and their ratio: the free series  have discrete predictive distributions, while the constrained series has a continuous one. 
Extending reconciliation via projection to such settings is challenging. The  method of \citet{ZHANG2024143} for discrete forecasts does not scale to large problems, and
there are so far no applications of
projection-based reconciliation to mixed-type predictive distributions.

Reconciliation via conditioning has already been applied to discrete \citep{CORANI2024457,zambon2024efficient} and mixed \citep{pmlr-v244-zambon24a} predictive distributions, and therefore appears more promising to manage such cases also under nonlinear constraints. However, our current UKF implementation assumes Gaussian predictive distributions for the free and constrained series and approximates the reconciled distribution of the free series by a Gaussian. These approximations are not well suited to discrete predictive distributions and may also be inaccurate under strong nonlinearities. Moreover, the assumption of independence between the free and constrained base forecasts (Eq.~\eqref{eq:rec_distr_cond_product}) may be unrealistic in some applications.

\section{Experiments}

We compare nonlinear reconciliation via \textit{projection} and via \textit{conditioning} using the following datasets:

\begin{enumerate}
    \item synthetic data with different nonlinear constraints;
    \item Swiss data on immigration and citizenship rates;
    \item  Australian tourism data, including the ratio of tourists in each region to the total number of tourists in the country.
\end{enumerate}

In our experiments, we evaluate all the results at the forecast horizon $h=1$. The code is available in the \textit{Github repository}\footnote{\url{https://github.com/ijfpnlr-svg/ijf_sub}} and allows reproducing our experiments. The  computation times were obtained on a machine with Intel Core i7-1370P processor (20 CPU cores) and 32 GB of RAM, running \textit{Linux}.

\paragraph{Probabilistic bottom-up}

As a benchmark method we consider a nonlinear \textit{probabilistic bottom-up} (\textbf{PBU}), which is the probabilistic counterpart of the nonlinear bottom-up used by \citet{girolimetto2025forecastreconciliationnonlinearconstraints} for point forecasts. Given samples $\bH^{(i)}$, for $i=1,\dots,M$, drawn from the  base forecast distribution of the free time series, PBU obtains the reconciled distribution by applying the FTA map $f$ to each sample:
\begin{align*}
    \yT^{(i)}_{PBU} = f(\bH^{(i)}), \quad i = 1, \dots, M.
\end{align*}

\paragraph{Evaluation metrics}

We use the \textit{energy score} \citep{GneitingRaftery2007} to evaluate the joint predictive distributions. At the evaluation time step $s$, we compute the sample approximation of the energy score given the actual observations $\ybf \in \mathbb{R}^n$ and $M$ samples $\xbf^{(j)}$ from the predictive distribution as
\begin{equation*}
\text{ES}_{s}
=
\frac{1}{M} \sum_{j=1}^M \| \xbf^{(j)} - \ybf \|_2 
- \frac{1}{2M^2}
\sum_{j=1}^M \sum_{k=1}^M \| \xbf^{(j)} - \xbf^{(k)} \|_2, 
\end{equation*}
where $\|.\|_2$ denotes the Euclidean norm. For the experiments we consider the relative Energy score:

\begin{equation*}
    \text{RelES}_{method} = \frac{\frac{1}{T_s}
    \sum_{s=1}^{T_s}
    \text{ES}_{s,method}}{\frac{1}{T_s}
    \sum_{s=1}^{T_s}\text{ES}_{s,base}},
\end{equation*}
where $T_s$ is the total number of evaluation time steps.

\begin{figure}[h!]
    \centering
    \includegraphics[width=0.9\linewidth]{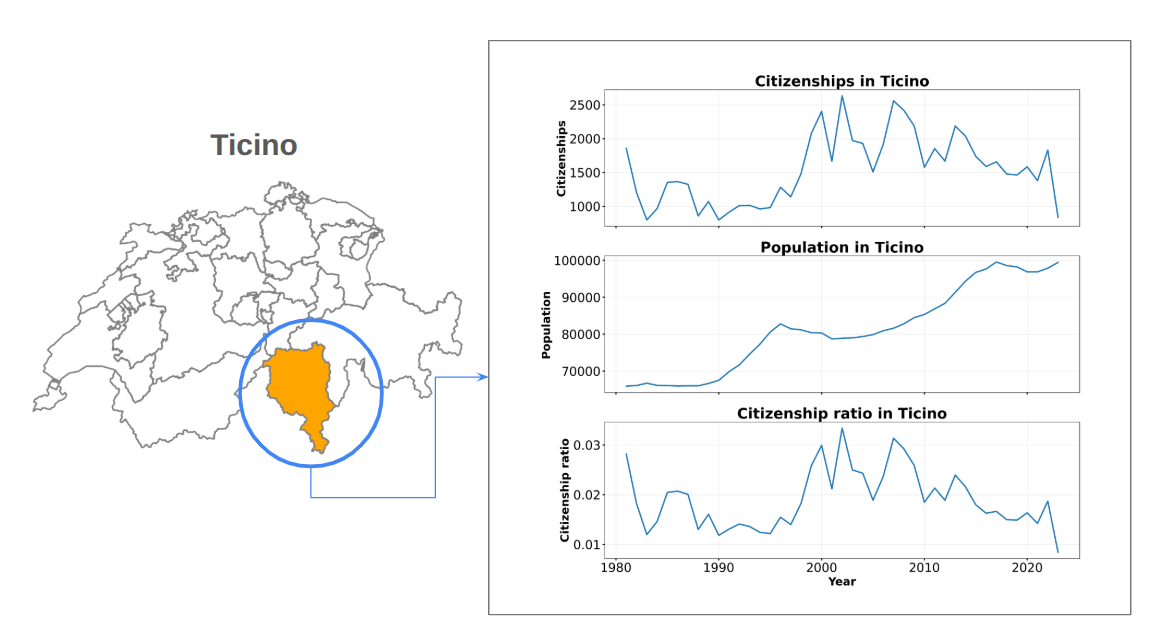}
    \caption{Number of acquired citizenships per year, existing foreigners and the citizenship ratio for Canton Ticino.}
    \label{fig:TI}
\end{figure}

The Energy Score is generally dominated by the time series with the largest scales. We therefore report the Energy Score for the simulated data, where the scales of the time series can be controlled. In contrast, the real data sets contain time series with large differences in scale (see \figref{fig:TI}). For these case studies, we evaluate forecast accuracy using the CRPS \citep{GneitingRaftery2007} (the univariate version of Energy score) on each time series. For completeness, we also report it in the simulation . At the evaluation time step $s$, we compute the sample approximation of the CRPS given the actual observations $y \in \rr$ and $M$ samples $x^{(j)}$ from the predictive distribution as
\begin{equation*}
\text{CRPS}_{s}
=
\frac{1}{M} \sum_{j=1}^M | x^{(j)} - y | 
- \frac{1}{2M^2}
\sum_{j=1}^M \sum_{k=1}^M | x^{(j)} - x^{(k)}|. 
\end{equation*}
In particular we report the geometric mean of the relative CRPS \citep{GIROLIMETTO20241134} across time series:
\begin{equation*}
    \mathrm{RelCRPS}_{\text{method}}
    =
    \left(
    \prod_{i=1}^{n}
    \frac{\frac{1}{T_s}
    \sum_{s=1}^{T_s}
    \text{CRPS}_{s,method,i}}{\frac{1}{T_s}
    \sum_{s=1}^{T_s}\text{CRPS}_{s,base,i}}
    \right)^{1/n},
\end{equation*}
where $T_s$ is the total number of evaluation time steps, and the additional subscript $i$ is used for the $i^{th}$ time series, since CRPS is an univariate score.

\paragraph{Significance tests}
To assess statistical uncertainty around the reported relative scores, we use a block-bootstrap procedure similar to the Model Confidence Set (MCS) \citep{HansenLundeNason2011}, which resamples rolling forecast origins in blocks, recomputes the full relative-score statistic at each bootstrap replication, and sequentially eliminates methods that are significantly worse. We describe our procedure in \ref{app:sig}.

\subsection{Synthetic data}

\subsubsection{Data generation and base forecasts}

\begin{figure}
    \centering
    \includegraphics[width=1\linewidth]{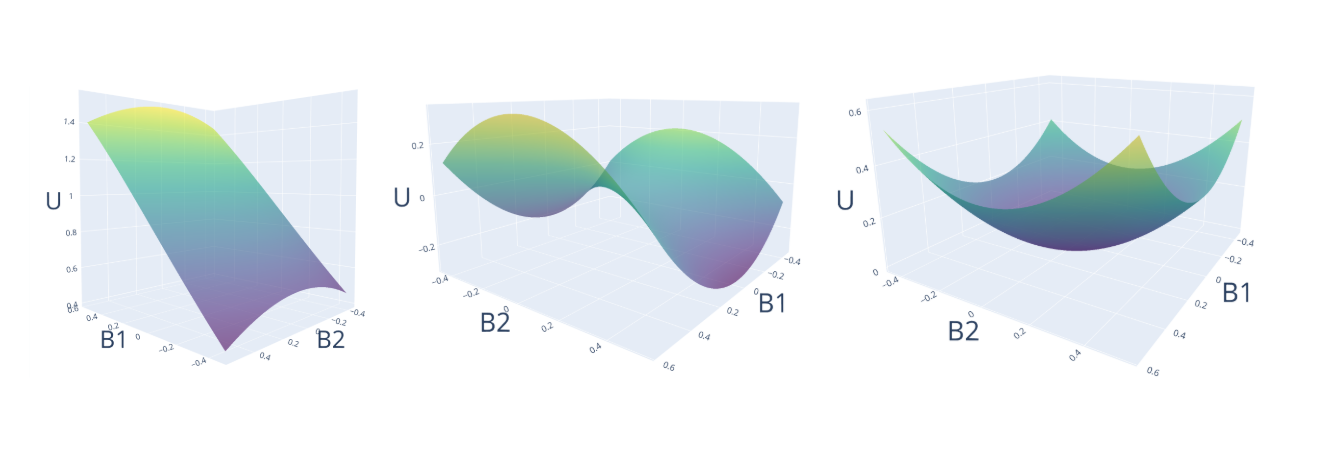}
    \caption{Simulation study: 3-D surfaces generated by the synthetic data - ripples (left), saddle (middle), and paraboloid (right).
    }
    \label{fig:sim}
\end{figure}

We compare the reconciliation methods on different nonlinear surfaces in $\mathbb{R}^3$. We generate two free-level time series $B_{1,t}$ and $B_{2,t}$ as $AR(1)$ processes:
\begin{align*}
    B_{1,t} = \phi_1B_{1,t-1}+\epsilon_{1,t}, \\
    B_{2,t} = \phi_2B_{2,t-1}+\epsilon_{2,t},
\end{align*}
where $\epsilon_{1,t} \sim \mathcal{N}(0,0.1^2)$ and $\epsilon_{2,t} \sim \mathcal{N}(0,0.1^2)$. We set $\phi_1 = \phi_2 = 0.9$ for all the surfaces and we generate data for $1000$ time steps. We use this  structure for the free level time series $B_t = [B_{1,t}, B_{2,t}]^T$ and generate the constrained level time series for the surfaces \textit{paraboloid}, \textit{saddle}, and \textit{ripples}, using the equations given in \tabref{tab:nlr_fun_sim}. \figref{fig:sim} shows the surfaces in the 3-D Euclidean space. 

\begin{table}[h!]
    \centering
\begin{tabular}{lccc}
    \toprule
    \textit{Surface} & \textbf{paraboloid} & \textbf{saddle} & \textbf{ripples} \\
    \midrule
    \textit{Constraint} & $\quad U_t = B_{1,t}^2+B_{2,t}^2 \quad$ & $\quad U_t = B_{1,t}^2-B_{2,t}^2 \quad$ & $\quad U_t = \sin B_{1,t} + \cos B_{2,t} \quad$ \\
    \bottomrule
\end{tabular}
    \caption{Nonlinear surfaces of the simulation study.}
    \label{tab:nlr_fun_sim}
\end{table}

Each time series is split into training and test sets using an 80\%–20\% chronological partition. Base forecasts are constructed separately for each time series by fitting a random forest  on the training data; the fitted model is then used to generate one-step-ahead forecasts across the test period using the actual lagged observations, without re-estimating the model. We use a random forest regressor with lag 1 for forecasting each series. To obtain probabilistic base forecasts, we apply a joint residual bootstrap based on the in-sample residual vectors of the three series. We jointly resample the residuals as done by \citet{PANAGIOTELIS2023693} to preserve the contemporaneous dependence structure across $U_{t+1}$, $B_{1,t+1}$, and $B_{2,t+1}$.

\subsubsection{Results}

\begin{table}[h!]
    \centering
    \begin{tabular}{llcccccc}
    \toprule
    \multicolumn{2}{c}{\multirow{2}{*}{\textit{Methods}}} & \multicolumn{3}{c}{\textit{RelCRPS}} & \multicolumn{3}{c}{\textit{RelES}} \\
    \cmidrule(l){3-5} \cmidrule(l){6-8}
    & & \textbf{paraboloid} & \textbf{saddle} & \textbf{ripples} & \textbf{paraboloid} & \textbf{saddle} & \textbf{ripples} \\
    \midrule
    \multirow{2}{*}{Baseline} & Base & 1.00 & 1.00 & 1.00 & 1.00 & 1.00 & 1.00 \\
     & PBU & 1.02 & 0.99 & 1.00 & 0.99 & 0.97 & 0.97 \\
    \midrule
    \multirow{3}{*}{Projection} & OLS & 0.98 & 0.98 & 0.96 & 0.96 & 0.96 & 0.93 \\
     & WLS & 0.97 & 0.99 & 0.95 & 0.95 & 0.97 & 0.93 \\
     & FULL & 0.97 & 0.99 & 0.95 & 0.95 & 0.97 & \textbf{0.92} \\
    \midrule
    \textit{Conditioning} & UKF & \textbf{0.94} & \textbf{0.96} & \textbf{0.94} & \textbf{0.93} & \textbf{0.95} & \textbf{0.92} \\
    \bottomrule
    \end{tabular}
    \caption{Relative CRPS and Energy Score for the simulation study. The best method is highlighted in bold.}
    \label{tab:simulation_study}
\end{table}

\tabref{tab:simulation_study} summarizes the results averaging over the different splits of the data. Probabilistic reconciliation by both via projection and conditioning improves the Relative Energy Score and Relative CRPS compared to the base forecast and the probabilistic bottom-up. However, the UKF is the most accurate approach on all metrics.
Among projections, the \textit{WLS} generally works better than \textit{OLS} as already noted in linear reconciliation \citep{ATHANASOPOULOS2024430, Wickramasuriya03042019}. There is no improvement from WLS to {FULL} possibly because the free time series were independently generated. 

%
The UKF is always retained in the final model confidence test together with some of the projection-based methods (\tabref{tab:sim_mcs_membership}). A detailed record of the sequential MCS elimination steps is reported in \textit{Appendix~S2} of the supplementary material.
Using 2,000 samples, the UKF is generally three orders of magnitude faster than projection across all surfaces (\tabref{tab:runtime}).


\begin{table}[h!]
    \centering
    \begin{tabular}{ccc}
    \toprule
     Paraboloid & Saddle & Ripples \\
    \midrule
     \color{brown}{\textbf{UKF}} & \color{brown}{\textbf{UKF}} & \color{brown}{\textbf{UKF}} \\
     \color{brown}{\textbf{OLS}} & \color{brown}{\textbf{OLS}} & \color{brown}{\textbf{FULL}} \\
     PBU & \color{brown}{\textbf{PBU}} & \color{brown}{\textbf{WLS}} \\
     FULL & FULL & \color{brown}{\textbf{OLS}} \\
     WLS & Base & \color{brown}{\textbf{PBU}} \\
     Base & WLS & Base \\
    \bottomrule
    \end{tabular}
    \caption{Outcome of the sequential studentized MCS procedure for the simulation study at \(\alpha=0.05\). Bold brown entries indicate the methods retained in the final model confidence set. Eliminated methods are ordered according to the sequential MCS elimination path, with the first eliminated method shown at the bottom.}
    \label{tab:sim_mcs_membership}
\end{table}

\begin{table}[h!]
    \centering
    \begin{tabular}{cccc}
    \toprule
     & \multicolumn{3}{c}{\textit{Computation time} (seconds)} \\
    \cmidrule(lr){2-4}
    \textit{Method} & \textit{paraboloid} & \textit{saddle} & \textit{ripples} \\
    \midrule       
    Projection (FULL)  & $1.83$ &  $1.10$ & $1.15$ \\
    UKF  & $1.6\times10^{-3}$ &  $1.6\times10^{-3}$ &  $2.2\times10^{-3}$ \\
    \bottomrule
    \end{tabular}
    \caption{Average runtime of reconciliation for different surfaces.}
    \label{tab:runtime}
\end{table}

\subsubsection{Scalability with high dimension}
The sigma-point approximation of the UKF may become numerically unstable when dealing with high-dimensional states \citep{Wu2006NumericalIntegration}; in our setting, the state dimension corresponds to the number of free time series $n_b$.
We therefore assess the accuracy of the UKF and the other methods increasing $n_b$ to $10$, $20$, and $50$.
We thus consider the high dimensional paraboloid, saddle and ripple as follows :
\begin{align}
    U_t &= B_{1,t}^2 + \dots + B_{n_b,t}^2 \qquad &\text{(paraboloid),}
    \label{eq:high_dim_cons1}\\
    U_t &= B_{1,t}^2 - \dots - B_{n_b,t}^2 \qquad &\text{(saddle),} 
    \label{eq:high_dim_cons2}\\
    U_t &= \sin B_{1,t} + \dots + \cos B_{n_b,t} \qquad &\text{(ripple).} 
    \label{eq:high_dim_cons3}
\end{align}
The results are provided in \tabref{tab:ablation_results} of \ref{app:sim_high}; in general  as $n_b$ increases, the gains from reconciliation shrink and the differences between methods narrow. This might be also due by the fact that as $n_b$ increases, the problem becomes increasingly imbalanced, with many free time series reconciled against a single nonlinear constraint.
\tabref{tab:sim_mcs_membership} shows that even in higher dimension UKF is retained in the final MCS set except for the single case of paraboloid at $n_b=50$. A detailed account of the elimination steps can be found in \textit{Appendix~S3} of the supplementary material.  

\subsubsection{Complexity analysis for computation time}

\begin{figure}[h!]
     \centering
     \includegraphics[width=0.7\linewidth]{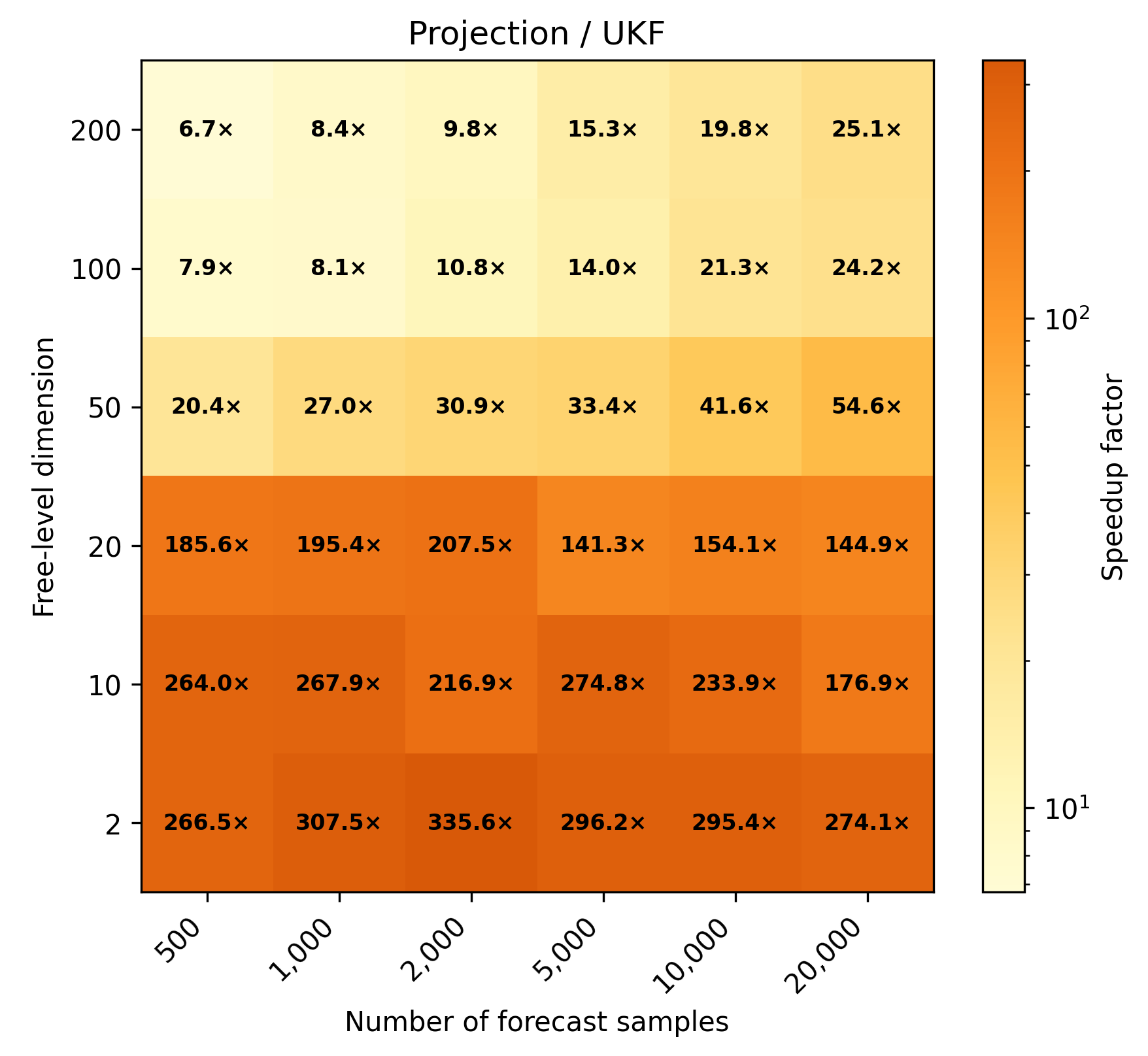}
     \caption{Computation time speedup for UKF compared to projection}
     \label{fig:comp_time}
\end{figure}

We do further complexity analysis of the computation times based on different sample sizes and dimensions of the free time series. We consider the sample sizes 500, 1000, 2000, 5000, 10000, 20000 and dimensions of free time series as 2, 10, 20, 50, 100, 200. The constraints are chosen as in \eqtref{eq:high_dim_cons1}, \eqtref{eq:high_dim_cons2}, and \eqtref{eq:high_dim_cons3}. Detailed runtime trends for the two reconciliation strategies are reported in \ref{app:comp_time}. In brief, the projection-based method shows a strong dependence on the number of forecast samples (since it needs to project each sample) and on the number of free time series (as the computational complexity to perform the numerical optimization grows in high dimensions). The UKF-based conditioning method is mostly driven by the dimension of the free time series (to compute Sigma points that grows in high dimensions) and is only weakly affected by the number of samples (for the final sampling step). A summarized comparison of the computational speedup defined as the ratio of the computation time taken by projection to the computation time taken by UKF is shown in \figref{fig:comp_time} as the average over the three surfaces. The UKF-based reconciliation becomes increasingly advantageous as the number of forecast samples grows. This effect is particularly significant when the number of free time series is small, where the speedup reaches very large values. Although the speedup decreases as the number of free time series increases, the smallest speedup observed in our experiments is still approximately 6-7 times.

\subsection{Empirical studies}

\subsubsection{Datasets}

\paragraph{Swiss demographic ratios}

We now consider forecasting two socio-demographic rates in Switzerland: immigration rate and acquired citizenship rate.
Switzerland is subdivided into 26 Cantons.
We extract, for each Canton and for Switzerland, annual counts of foreign  population,  immigration counts, and  acquired Swiss citizenships  from  1981 to 2024. These are the components required to construct the immigration-to-population ratio and the citizenship-to-population ratio. We obtain the data from the \textit{Swiss Federal Statistical Office (FSO)}\footnote{\url{https://www.pxweb.bfs.admin.ch/api/v1/en/px-x-0102020000_104/px-x-0102020000_104.px}}.

\begin{figure}[h!]
\begin{tikzpicture}[scale=0.9, 
    every node/.style={rectangle, draw, align=center, rounded corners=4, font=\large, minimum height=0.65cm, minimum width=0.9cm},
    topbox/.style={fill=green!10},
    bottombox/.style={fill=orange!10},
    fork down/.style={
      to path={
        let \p1 = (\tikztostart.south),
            \p2 = (\tikztotarget.north),
            \n1 = {(\y1+\y2)/2} in
        (\p1) -- (\x1,\n1) -- (\x2,\n1) -- (\p2) \tikztonodes
      }
    }
]

\node[topbox] (I) at (-6,3) {$I^{CH}$};
\node[topbox] (R) at (0,2) {$R^{CH}$};
\node[topbox] (P) at (6,3) {$P^{CH}$};

\node[bottombox] (IA) at (-8,0) {$I^{AG}$};
\node[bottombox, right=1cm of IA] (dots2) {$\cdots$};
\node[bottombox] (IZ) at (-4,0) {$I^{ZH}$};
 
\node[topbox] (RA) at (-2,1) {$R^{AG}$};
\node[topbox, right=1cm of RA] (dots2) {$\cdots$};
\node[topbox] (RZ) at (2,1) {$R^{ZH}$};
 
\node[bottombox] (PA) at (4,0) {$P^{AG}$};
\node[bottombox, right=1cm of PA] (dots2) {$\cdots$};
\node[bottombox] (PZ) at (8,0) {$P^{ZH}$};
 
\path[fork down, thick] (P) edge (PA)
      (P) edge (PZ);
\path[fork down, thick] (I) edge (IA)
      (I) edge (IZ);
      
\draw[dashed, very thick] (P) -- (R);
\draw[dashed, very thick] (I) -- (R);
\draw[dashed, very thick] (PA) -- (RA);
\draw[dashed, very thick] (IA) -- (RA);
\draw[dashed, very thick] (PZ) -- (RZ);
\draw[dashed, very thick] (IZ) -- (RZ);

\end{tikzpicture}
   \caption{The structure of immigration rates 
   disaggregated by Swiss Cantons contains both linear (solid) and nonlinear (dashed) constraints. The orange nodes represent the free time series, and the green nodes represent  the constrained time series.
   \label{fig: imm_rate}}
\end{figure}
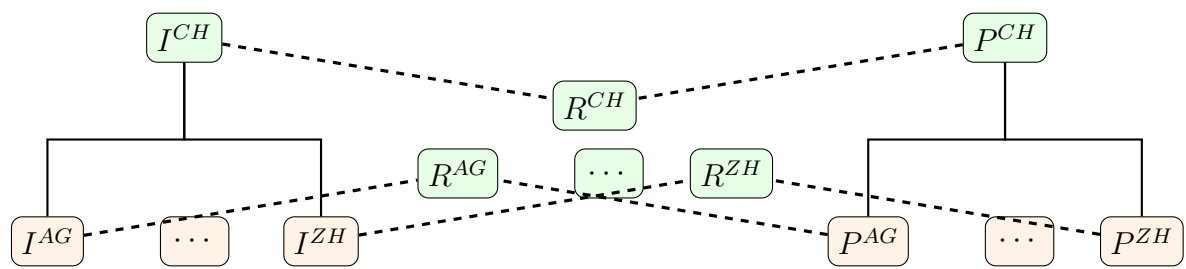

The structure of the constraints for the immigration rate is shown in \figref{fig: imm_rate}. At the national level, the number of immigrants $I^{CH}$, the foreign population $P^{CH}$, and the immigration rate $R^{CH}$ are linked by the nonlinear constraint
\[
R^{CH} = \frac{I^{CH}}{P^{CH}},
\]
represented by dashed lines. The same nonlinear constraint holds for each of the 26 Cantons, with Aargau (AG) and Zürich (ZH) shown as examples:
\[
R^{AG} = \frac{I^{AG}}{P^{AG}},
\qquad
R^{ZH} = \frac{I^{ZH}}{P^{ZH}}.
\]
Some constraints are linear: the immigration counts of the 26 Cantons sum to the national immigration count, and the cantonal foreign populations sum to the national foreign population. These constraints are represented by solid lines.
The same data structure is used for forecasting the acquired citizenship rate.

We compute the base probabilistic forecasts for each time series using \texttt{auto.arima()} from the R package \texttt{forecast} \citep{hyndman2008automatic}. For each time series, we produce forecasts using a fixed-length training window of $12$ years. We adopt rolling windows that advance by one year at a time and we have $30$ rolling windows in total. We produce forecasts for the horizon $h=1$, i.e., for the next one year. We draw $M=1000$ samples out of the base forecast distribution, which we reconcile using the two approaches mentioned before.

\paragraph{Australian tourism ratio}

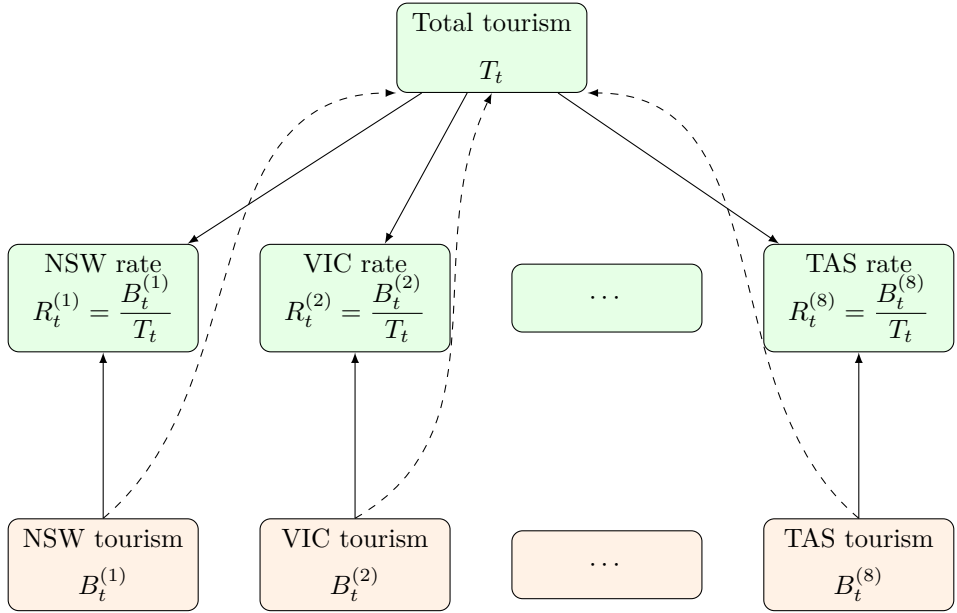
\begin{figure}[t]
\centering
\begin{tikzpicture}[
    >=latex,
    node distance=1.6cm and 1.4cm,
    every node/.style={font=\small},
    box/.style={
        draw,
        rounded corners,
        align=center,
        minimum width=2.5cm,
        minimum height=0.9cm
    },
    topbox/.style={
        box,
        fill=green!10
    },
    ratiobox/.style={
        box,
        fill=green!10
    },
    bottombox/.style={
        box,
        fill=orange!10
    }
]

\node[topbox, xshift=-2cm] (total) {Total tourism\\$T_t$};

\node[ratiobox, below left=2cm and 2.6cm of total] (r1) {NSW rate\\$R_t^{(1)}=\dfrac{B_t^{(1)}}{T_t}$};
\node[ratiobox, right=0.8cm of r1] (r2) {VIC rate\\$R_t^{(2)}=\dfrac{B_t^{(2)}}{T_t}$};
\node[ratiobox, right=0.8cm of r2] (dots1) {$\cdots$};
\node[ratiobox, right=0.8cm of dots1] (r8) {TAS rate\\$R_t^{(8)}=\dfrac{B_t^{(8)}}{T_t}$};

\node[bottombox, below=2.2cm of r1] (b1) {NSW tourism\\$B_t^{(1)}$};
\node[bottombox, right=0.8cm of b1] (b2) {VIC tourism\\$B_t^{(2)}$};
\node[bottombox, right=0.8cm of b2] (dots2) {$\cdots$};
\node[bottombox, right=0.8cm of dots2] (b8) {TAS tourism\\$B_t^{(8)}$};

\draw[->] (total) -- (r1);
\draw[->] (total) -- (r2);
\draw[->] (total) -- (r8);

\draw[->] (b1) -- (r1);
\draw[->] (b2) -- (r2);
\draw[->] (b8) -- (r8);

\draw[->, dashed] (b1.north) to[out=40,in=180] (total.south west);
\draw[->, dashed] (b2.north) to[out=30,in=245] (total.south);
\draw[->, dashed] (b8.north) to[out=140,in=360] (total.south east);

\end{tikzpicture}
\caption{Structure of the Australian tourism case study. The bottom level contains quarterly tourism counts for each state, the top level contains the national total, and an additional nonlinear level is defined through the state tourism rates $R_t^{(i)} = B_t^{(i)}/T_t$.}
\label{fig:tourism_hierarchy}
\end{figure}

We also create a set of time series with nonlinear constraints on the Australian Tourism dataset available in the R package \texttt{tsibble} \citep{tsibbledata}. We consider the portion of the dataset which contains the quarterly tourism counts from the 8 different states of Australia. The data comprises observations over 20 years (80 quarters) from the first quarter of 1998 to the last quarter of 2017. \figref{fig:tourism_hierarchy} shows the structure of the data, where the free and constrained time series are color coded as light orange and light green respectively. We create a new time series (tourism rate) as:

\begin{equation}\label{eq:aus_tour}
        \text{Rate}_{state,t} = \frac{\# \text{Tourism}_{state,t}}{\#\text{Tourism}_{AUS,t}}
\end{equation}

Similar to the Swiss datasets, we generate base forecasts with \texttt{auto.arima} \citep{hyndman2008automatic} using 40 rolling windows and a forecast horizon $h=1$.

\subsubsection{Results}

\begin{table}[h!]
    \centering
\begin{tabular}{ccccc}
    \toprule
    & \multirow{2}{*}{Method} & \multicolumn{3}{c}{Dataset} \\
    &  & \textit{Immigration} & \textit{Citizenship} & \textit{Australian Tourism} \\
    \midrule
    \multirow{2}{*}{Baseline}
    & Base & 1 & 1 & 1 \\
    & PBU & 0.95 & 1 & 1.03 \\
    \midrule
    \multirow{3}{*}{Projection} 
    & OLS & 1.21 & 1.36 & 1.09 \\
    & WLS & 0.94 & 0.99 & 1.02 \\
    & FULL & 0.93 & 0.97 & 1.02 \\
    \midrule
    Conditioning
    & UKF & \textbf{0.91} & \textbf{0.96} & \textbf{0.97} \\
    \bottomrule
\end{tabular}
    \caption{RelCRPS compared to base forecast for the Swiss demographic datasets and the Australian tourism dataset}
    \label{tab:empirical_relcrps}
\end{table}
\tabref{tab:empirical_relcrps} summarizes the results. For the Swiss datasets, all the reconciliation methods other than the OLS perform better than the base forecast and the probabilistic bottom-up. {UKF} is the best performing reconciliation approach, improving the base forecast by $9\%$ and $4\%$. 
As mentioned earlier, both datasets contain time series with large differences in scale, which are also reflected in the base forecast errors. OLS does not account for these differences, which explains its poor performance.
In both case studies, the projection methods benefit from increasingly accurate modeling of the covariance structure, moving from OLS to WLS to FULL. A similar pattern has also been noted in nonlinear reconciliation of point forecasts \citep{nespoli2026nonlinear, girolimetto2025forecastreconciliationnonlinearconstraints}.

These results are also in-line with the simulation study, which suggested conditioning-based reconciliation methods work as accurately or sometimes even better than the projection approach. Similar results can be observed in the nonlinear level (Cantonal ratio) of the two datasets reported in the \textit{Appendix~S4.1} of the supplementary material.

\figref{fig:crps_cant} shows a canton-wise summary of the relative CRPS ratio for the immigration and citizenship datasets, comparing UKF to the two baselines. Green shading indicates an improvement over the baseline (darker = better), while red shading indicates a deterioration (darker = worse). For the immigration data, UKF improves over the base forecast in every canton. For the citizenship data, UKF improves over the base forecast in most cantons, and where it does not, the degree of deterioration is much smaller than the improvements observed elsewhere. UKF also generally improves over PBU across both the datasets.

\begin{figure}[htbp]
    \centering
    \includegraphics[width=0.7\textwidth]{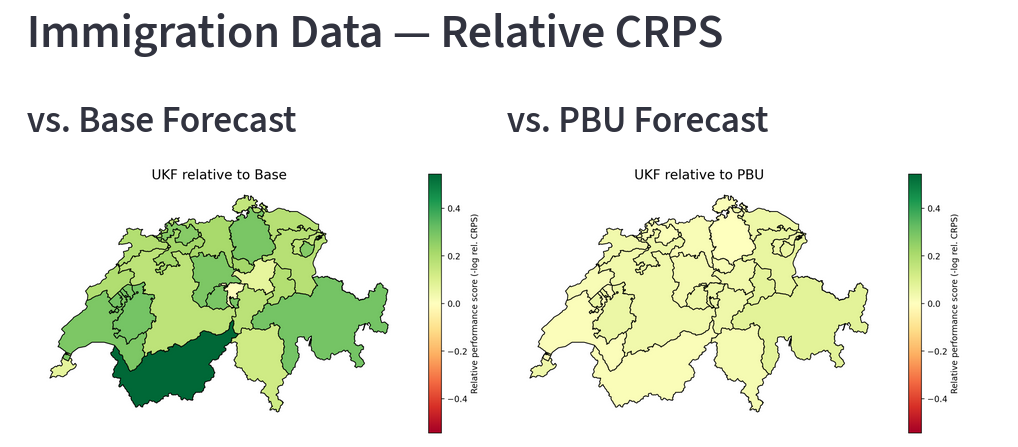}

    \vspace{0.5cm}

    \includegraphics[width=0.7\textwidth]{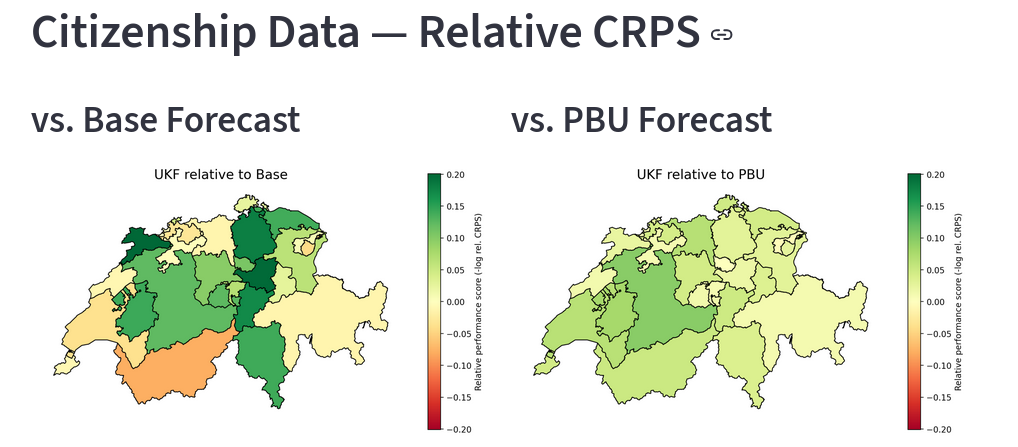}

    \caption{Improvement for UKF against the Base forecast and PBU for each Canton on the two datasets.
    }
    \label{fig:crps_cant}
\end{figure}

For the Australian tourism dataset, we observe a similar trend - reconciliation via conditioning provides a better score against the base forecast. The UKF achieves an improvement of $2.6\%$ over the base forecast, and it turns out to be the only reconciliation method that improves the score against the base forecast. Although the projection-based methods (other than the OLS due to scaling issue) achieves a better score than the PBU in general, they do not improve on the base forecast.

\subsubsection{MCS tests}

\begin{table}[h!]
    \centering
    \begin{tabular}{ccc}
    \toprule
     Immigration & Citizenship & Australian Tourism \\
    \midrule
     \color{brown}{\textbf{UKF}} & \color{brown}{\textbf{UKF}} & \color{brown}{\textbf{UKF}} \\
     PBU & \color{brown}{\textbf{FULL}} & \color{brown}{\textbf{Base}} \\
     FULL & WLS & \color{brown}{\textbf{FULL}} \\
     WLS & PBU & \color{brown}{\textbf{WLS}} \\
     Base & Base & \color{brown}{\textbf{PBU}} \\
     OLS & OLS & OLS \\
    \bottomrule
    \end{tabular}
    \caption{Outcome of the sequential studentized MCS procedure for the Swiss demographic datasets and the Australian tourism dataset (level full) at \(\alpha=0.05\). Bold brown entries indicate the methods retained in the final model confidence set. Eliminated methods are ordered according to the sequential MCS elimination path, with the first eliminated method shown at the bottom.}
    \label{tab:empirical_mcs_membership}
\end{table}

\tabref{tab:empirical_mcs_membership} reports the methods retained in the final studentized MCS for the two Swiss datasets and the Australian tourism dataset. At the \(5\%\) significance level, for the immigration dataset, UKF is the only method retained in the final MCS. This provides evidence that the remaining methods are significantly inferior to UKF according to the relative CRPS criterion. For the citizenship dataset, the final MCS contains both UKF and \textit{FULL}. Hence, while the other methods are eliminated, FULL cannot be statistically distinguished from UKF on the basis of the available evidence.

The MCS significance analysis for the Australian tourism dataset shows that, at the \(\alpha = 0.05\) level, OLS is eliminated from the final MCS, while UKF and the remaining reconciliation methods are retained. Thus, there is evidence that OLS is significantly inferior according to the relative CRPS criterion, but the available evidence is not sufficient to statistically distinguish UKF from the other retained methods. A detailed elimination record for the MCS tests are reported in \textit{Appendix~S4.2} of the supplementary material.

\subsubsection{Computation times}

\begin{table}[h!]
    \centering
    \begin{tabular}{cccc}
    \toprule
    & \multicolumn{3}{c}{\textit{Computation time} (seconds)} \\
    \cmidrule(lr){2-4}
    \textit{Method} & \textit{Immigration} & \textit{Citizenship} & \textit{Australian Tourism} \\
    \midrule       
    Projection (FULL)  & $5.19$ &  $4.07$ & $2.46$ \\
    UKF  & $6.3\times10^{-2}$ &  $6.7\times10^{-2}$ & $2.8\times10^{-3}$ \\
    \bottomrule
    \end{tabular}
    \caption{Average runtime of reconciliation for the Swiss datasets and the Australian tourism dataset computed on a single time step.}
    \label{tab:empirical_runtime}
\end{table}

\tabref{tab:empirical_runtime} shows the average runtime (in seconds) for the two approaches of reconciliation done with 1000 sample points. For the Swiss datasets, the UKF has a runtime which is faster than projection (FULL) by approximately 2 orders of magnitude. For the Australian tourism dataset, the UKF reconciles faster than the projection (FULL) by approximately 3 orders of magnitude, for a sample size of 1000.

\section{Conclusions}

We studied nonlinear probabilistic forecast reconciliation for the first time, extending  both  \textit{reconciliation via projection} and  \textit{reconciliation via conditioning} to the nonlinear setting. In our experiments, reconciliation in general improves the accuracy compared to both the base forecast and probabilistic bottom-up. The UKF-based reconciliation via conditioning outperforms, slightly but consistently, any other method in terms of probabilistic scores, while being substantially more computationally efficient than projection.

However, these results are still empirical, and theoretical guarantees for nonlinear reconciliation, such as reductions in Euclidean distance-based losses under local curvature conditions, are currently limited to point forecasts \citep{nespoli2026nonlinear, girolimetto2025forecastreconciliationnonlinearconstraints}. Establishing analogous results for probabilistic reconciliation remains an important open problem.

Both proposed reconciliation approaches admit natural extensions that we leave to future work. 
For conditioning, the reconciled distribution in Eq.~\eqref{eq:rec_distr_cond_bottom} is defined
for arbitrary base forecasts; moving beyond the Gaussian approximation would require replacing the UKF with sample-based approximations, for example via importance sampling or MCMC. 
For projection, an alternative to the nearest-point map is to estimate the reconciliation map $\psi$ by directly minimizing a proper scoring rule, such as the Energy Score or the Variogram Score, as done by \citet{PANAGIOTELIS2023693} in the linear case. 
To do so, $\psi$ could be defined, for instance, as a composition $\psi = f\circ g$, where $g:\rr^n\to\rr^{n_b}$ is 
a parametrized nonlinear map (e.g., a neural network), so that $\psi$ maps onto $\mm$ by construction.

Finally, nonlinear probabilistic reconciliation raises the question of how to define a suitable point forecast from the reconciled distribution. In the linear setting, the mean of a coherent distribution is itself coherent. Under nonlinear constraints, instead, the mean of a distribution supported on the coherent manifold generally does not lie on the manifold and is therefore incoherent. Alternative point summaries, such as the Fréchet mean computed on the coherent manifold \citep{frechet1948elements, karcher1977riemannian} defined with respect to geodesic distance on the manifold, may be considered as an alternative \citep{nespoli2026nonlinear}.


\section*{Acknowledgements}
Research funded by the Swiss National Science Foundation (grant $200021\_212164$), the European Union (project: 101160720 — ENERGENIUS), and the Swiss State Secretariat for Education, Research and Innovation (SERI) in the context of the Horizon Europe research and innovation programme project DR-RISE (Grant Agreement No 101104154) and REEFLEX (Horizon Europe, Grant Agreement No. 101096192).

\appendix

\section{Supplementary material}

Supplementary material associated with this article can be found in the accompanying Supplementary Material file. It includes 
the detailed description of reconciliation via the UKF (Appendix~S1) and the detailed elimination records for the MCS tests for the simulation study (Appendix~S2), the high dimensional simulations (Appendix~S3), and the empirical studies (Appendix~S4).

\section{Well-posedness of the projection}\label{app:wellposed}

We discuss conditions under which the nearest-point projection $\psi$, and thus the push-forward $\nutil=\psi_{\#}\nuhat$, is well defined.
We recall the assumptions stated in Sect.~\ref{sec:rec_via_proj}: that $\mm$ is non-empty and closed, and that $\nuhat$ is absolutely continuous with respect to the Lebesgue measure on $\rr^n$.

\paragraph{Existence} 
Since $\Wbf\succ0$, the map $\xbf\mapsto\|\yH-\xbf\|_\Wbf$ is continuous and coercive, so if $\mm$ is closed its minimum over $\mm$ is
attained, i.e. the set of minimizers
$\Psi(\yH):=\argmin_{\xbf\in\mm}\|\yH-\xbf\|_\Wbf$ is non-empty for every $\yH$. 
From Eq.~\eqref{eq:manifold}, $\mm$ is the graph over $\mathscr{B}$ of the FTC map $f_u$.
Since the graph of a continuous map on a closed set is closed\footnote{
Consider a sequence $\{\ybf_n\} \subset \mm$ that converges to some $\ybf$, 
and denote by $\{\bbf_n\}$ the free components of $\{\ybf_n\}$, so that $\bbf_n \to \bbf$.
Since $\mathscr{B}$ is closed, $\bbf\in\mathscr{B}$; 
and by continuity of $f_u$, $\ybf = \big(\bbf,f_u(\bbf)\big)\in\mm$.
}, 
$\mm$ is closed whenever $f_u$ is continuous on $\mathscr{B}$ and $\mathscr{B}$ is closed (e.g., when $\mathscr{B}$ is compact or $\mathscr{B}=\rr^{n_b}$).

However, existence may fail when $\mathscr{B}$ is not closed: for example, if $\mathscr{B} = \{(x,y)\in\rr^2:\, y\neq0\}$ and $f_u(x,y) = x/y$, 
no nearest point exists for any $\yH$ on the line $\{(0,0,c):\,c\in\rr\}$.
In this case, it is enough to restrict $\mathscr{B}$ to a compact set bounded away from $y=0$:
\[
\mathscr{B} = \{(x,y):|x|\le x_{\max},\ \epsilon\le y\le y_{\max}\},
\]
for some small $\epsilon>0$ and some large $x_{\max}, y_{\max}$.

\paragraph{Uniqueness} 
This holds only almost everywhere, which is what the push-forward requires. 
Since $\mm$ is closed and nonempty, $d(\yH)=\min_{\xbf\in\mm}\|\yH-\xbf\|_\Wbf$ is finite for every $\yH \in \rr^n$ and $1$-Lipschitz on $\rr^n$; hence, by Rademacher's theorem
\citep[e.g.,][Ch.~3.1]{evans2015measure}, it is differentiable Lebesgue-almost everywhere.
Wherever $d$ is differentiable the nearest point is unique \citep[Ch.~3]{cannarsa2004semiconcave}, 
so the set $N$ of points admitting more than one
projection has Lebesgue measure zero, and by absolute continuity $\nuhat(N)=0$.

\paragraph{Measurability} 
Since $\nuhat(N)=0$, it is enough to specify $\psi$ on the set $\rr^n\setminus N$, where the projection is single-valued and continuous\footnote{
If $\yH_k\to\yH$ with $\yH_k,\yH\in\rr^n\setminus N$, the sequence of nearest points $\psi(\yH_k)$ is bounded, as
$\|\yH_k-\psi(\yH_k)\|_\Wbf=d(\yH_k)\le\|\yH_k-\psi(\yH)\|_\Wbf$; 
the limit of any subsequence lies in $\mm$ and attains $d(\yH)$ by continuity of $d$, hence equals $\psi(\yH)$. Thus
$\psi(\yH_k)\to\psi(\yH)$.
}, 
hence Borel measurable. 
Therefore $\nutil=\psi_{\#}\nuhat$ is a well-defined Borel probability measure on $\mm$.


\section{Derivation of the reconciled distribution via conditioning}
\label{app:derivation_rec_cond}

Let $\YH = \left[\UH',\, \BH'\right]'$ be a random vector distributed as the base forecast distribution $\nuhat$.
We specify the reconciled distribution $\nutil_B$ of the free time series as the distribution of
\begin{equation}
\BT \;\:\sim\;\; \BH\;\Big| \left(\UH - f_u(\BH) = \mathbf{0}\right).
\end{equation}
Let $\nuhat \in \mathcal{F}_{\mathbb{R}^n}$ be the base forecast distribution;
we assume that $\nuhat$ is absolutely continuous, and we denote by $\pihat$ its density. 
Following \citet{zambon2024efficient}, we define $\Zbf = \hat \Ubf - f_u(\hat \Bbf)$. The joint density function of $(\Zbf,\hat \Bbf)$ can be computed as:
\begin{equation*}
    \pi_{(\Zbf,\hat \Bbf)}(\zbf,\bbf) = \hat{\pi} (\zbf+f_u(\bbf),\bbf)
\end{equation*}
and the conditional density of $\BH\;|\;\Zbf=\mathbf{0}$ is the density of the reconciled forecast distribution of the free series:
\begin{align*}
    \pitil_B(\bbf) 
    &= \frac{\pi_{(\Zbf,\BH)}(\mathbf{0},\bbf)}{\int \pi_{(\Zbf,\BH)}(\mathbf{0},\xbf)\,d\xbf} \\
    &= \frac{\hat \pi(f_u(\bbf),\bbf)}{\int \hat \pi(f_u(\xbf),\xbf)\,d\xbf} \\
    &\propto \hat \pi(f_u(\bbf),\bbf).
\end{align*}

\section{Significance tests based on MCS}
\label{app:sig}

To test the significance of the relative scores reported in the experiments, we use a block-bootstrap procedure based on \citet{HansenLundeNason2011}. The procedure resamples rolling forecast origins in blocks, and for each bootstrap sample computes the relative CRPS statistic:
\begin{equation*}
    R_m^{*(b)}
    =
    \exp\left[
    \frac{1}{n}
    \sum_{j=1}^{n}
    \log\left(
    \frac{
    \frac{1}{T^*}\sum_{t\in\mathcal{I}_b}\mathrm{CRPS}_{m,j,t}
    }{
    \frac{1}{T^*}\sum_{t\in\mathcal{I}_b}\mathrm{CRPS}_{\mathrm{Base},j,t}
    }
    \right)
    \right],
\end{equation*}
where \(b=1,\dots,B\) indexes the bootstrap replication, \(m=1,\dots,M\) indexes the reconciliation method, \(\mathcal{I}_b\) is the set of resampled rolling origins in bootstrap replication \(b\), and \(T^*=|\mathcal{I}_b|\). Then we compute the bootstrap pairwise differences across methods,
\[
    D_{ij}^{*(b)} = R_i^{*(b)} - R_j^{*(b)},
\]
which are centered around the observed differences \(D_{ij}=R_i-R_j\), and studentized:
\begin{equation*}
    t_{ij}^{*(b)}
    =
    \frac{
    D_{ij}^{*(b)} - D_{ij}
    }{
    \hat{\sigma}_{ij}
    },
\end{equation*}
where \(\hat{\sigma}_{ij}\) is the bootstrap standard error of \(D_{ij}^{*(b)}\). For a current active set of methods \(\mathcal{M}\), we use the range statistic
\begin{equation*}
    T_{R,\mathcal{M}}^{*(b)}
    =
    \max_{i,j\in\mathcal{M},\,i\neq j}
    \left|
    t_{ij}^{*(b)}
    \right|.
\end{equation*}
The observed statistic \(T_{R,\mathcal{M}}\) is computed analogously from the observed studentized differences \(t_{ij}=D_{ij}/\hat{\sigma}_{ij}\). The null hypothesis of equal predictive ability within \(\mathcal{M}\) is rejected when \(T_{R,\mathcal{M}}\) exceeds the empirical \((1-\alpha)\)-quantile of the bootstrap distribution of \(T_{R,\mathcal{M}}^{*(b)}\). If the null is rejected, the worst method is removed according to
\begin{equation*}
    e_{\mathcal{M}}
    =
    \arg\max_{i\in\mathcal{M}}
    \sup_{j\in\mathcal{M}}
    t_{ij},
\end{equation*}
and the test is repeated on the reduced set. The final set contains the methods that cannot be statistically distinguished from the best method at significance level \(\alpha\).

\section{Scalability with dimension}
\label{app:sim_high}

For $n_b=10$, UKF is competitive with the projection methods on the paraboloid and saddle surfaces -- tying for the best CRPS on both and achieving the best Energy Score on the saddle surface -- but it is outperformed by WLS on the more nonlinear ripples surface. As $n_b$ increases to $20$ and $50$, UKF's relative advantage narrows further: it ties with the projection methods on the saddle surface, but is dominated by the base forecast on the paraboloid surface and by WLS/FULL on the ripples surface, particularly in terms of Energy Score at $n_b=50$, likely due to its known sensitivity to the stability of the Sigma points as the bottom-level dimension grows. \tabref{tab:sim_mcs_membership} shows that UKF is retained in the final MCS set across all the surfaces in high dimensions apart from paraboloid at $n_b=50$, which means although it marginally fails to be the best method in some cases, still it is not statistically worse than the best observed method.

\begin{table}[h!]
    \centering
    \scriptsize
    \resizebox{\textwidth}{!}{%
    \begin{tabular}{lllcccccc}
    \toprule
    \multirow{2}{*}{$n_b$} 
    & \multicolumn{2}{c}{\multirow{2}{*}{\textit{Methods}}} 
    & \multicolumn{3}{c}{\textit{CRPS}} 
    & \multicolumn{3}{c}{\textit{Energy Score}} \\
    \cmidrule(l){4-6} \cmidrule(l){7-9}
    & & 
    & \textbf{paraboloid} 
    & \textbf{saddle} 
    & \textbf{ripples} 
    & \textbf{paraboloid} 
    & \textbf{saddle} 
    & \textbf{ripples} \\
    \midrule
    \multirow{6}{*}{$10$}
    & \multirow{2}{*}{Baseline} 
        & Base  & \textbf{1.00} & 1.00 & 1.00 & 1.00 & 1.00 & 1.00 \\
    &   & PBU   & 1.01 & 1.02 & 1.01 & 1.01 & 1.01 & 1.01 \\
    \cmidrule(l){2-9}
    & \multirow{3}{*}{Projection} 
        & OLS   & \textbf{1.00} & \textbf{0.99} & 0.99 & \textbf{0.99} & 0.99 & 0.99 \\
    &   & WLS   & \textbf{1.00} & \textbf{0.99} & \textbf{0.98} & \textbf{0.99} & 0.99 & \textbf{0.98} \\
    &   & FULL  & \textbf{1.00} & \textbf{0.99} & 0.99 & \textbf{0.99} & 0.99 & \textbf{0.98} \\
    \cmidrule(l){2-9}
    & Conditioning 
        & UKF   & \textbf{1.00} & \textbf{0.99} & 0.99 
                & \textbf{0.99} & \textbf{0.98} & 0.99 \\
    \midrule
    \multirow{6}{*}{$20$}
    & \multirow{2}{*}{Baseline} 
        & Base  & \textbf{1.00} & 1.00 & 1.00 & \textbf{1.00} & 1.00 & 1.00 \\
    &   & PBU   & 1.02 & 1.01 & 1.02 & 1.03 & 1.01 & 1.04 \\
    \cmidrule(l){2-9}
    & \multirow{3}{*}{Projection} 
        & OLS   & 1.01 & \textbf{0.99} & \textbf{0.99} & \textbf{1.00} & \textbf{0.99} & \textbf{0.99} \\
    &   & WLS   & 1.01 & \textbf{0.99} & \textbf{0.99} & \textbf{1.00} & \textbf{0.99} & \textbf{0.99} \\
    &   & FULL  & 1.01 & \textbf{0.99} & \textbf{0.99} & \textbf{1.00} & \textbf{0.99} & \textbf{0.99} \\
    \cmidrule(l){2-9}
    & Conditioning 
        & UKF   & 1.01 & \textbf{0.99} & 1.00 
                & 1.01 & \textbf{0.99} & 1.00 \\
    \midrule
    \multirow{6}{*}{$50$}
    & \multirow{2}{*}{Baseline} 
        & Base  & \textbf{1.00} & \textbf{1.00} & \textbf{1.00} & \textbf{1.00} & \textbf{1.00} & 1.00 \\
    &   & PBU   & 1.02 & \textbf{1.00} & \textbf{1.00} & 1.05 & 1.00 & 1.00 \\
    \cmidrule(l){2-9}
    & \multirow{3}{*}{Projection} 
        & OLS   & 1.01 & \textbf{1.00} & \textbf{1.00} & \textbf{1.00} & \textbf{1.00} & 1.00 \\
    &   & WLS   & \textbf{1.00} & \textbf{1.00} & \textbf{0.99} & 1.01 & \textbf{0.99} & \textbf{0.97} \\
    &   & FULL  & \textbf{1.00} & \textbf{1.00} & \textbf{0.99} & 1.01 & \textbf{0.99} & \textbf{0.97} \\
    \cmidrule(l){2-9}
    & Conditioning 
        & UKF   & 1.01 & \textbf{0.99} & \textbf{0.99} 
                & 1.03 & \textbf{0.99} & 0.98 \\
    \bottomrule
    \end{tabular}
    }
    \caption{Relative CRPS and Energy Score for the simulation study for increasing numbers of free time series $n_b$. The best value within each $n_b$, score, and nonlinear constraint surface is highlighted in bold.}
    \label{tab:ablation_results}
\end{table}

\begin{table}[h!]
    \centering
    \scriptsize
    \resizebox{\textwidth}{!}{%
    \begin{tabular}{ccc|ccc|ccc}
    \toprule
    \multicolumn{3}{c}{$n_b=10$} & \multicolumn{3}{c}{$n_b=20$} & \multicolumn{3}{c}{$n_b=50$} \\
    \cmidrule(lr){1-3} \cmidrule(lr){4-6} \cmidrule(lr){7-9}
    Paraboloid & Saddle & Ripples & Paraboloid & Saddle & Ripples & Paraboloid & Saddle & Ripples \\
    \midrule
    \color{brown}{\textbf{BASE}} & \color{brown}{\textbf{UKF}} & \color{brown}{\textbf{WLS}}
    & \color{brown}{\textbf{BASE}} & \color{brown}{\textbf{UKF}} & \color{brown}{\textbf{OLS}}
    & \color{brown}{\textbf{BASE}} & \color{brown}{\textbf{UKF}} & \color{brown}{\textbf{UKF}} \\
    \color{brown}{\textbf{UKF}} & \color{brown}{\textbf{OLS}} & \color{brown}{\textbf{FULL}}
    & \color{brown}{\textbf{UKF}} & \color{brown}{\textbf{OLS}} & \color{brown}{\textbf{WLS}}
    & UKF & \color{brown}{\textbf{WLS}} & \color{brown}{\textbf{WLS}} \\
    \color{brown}{\textbf{WLS}} & FULL & \color{brown}{\textbf{UKF}}
    & PBU & WLS & \color{brown}{\textbf{UKF}}
    & FULL & \color{brown}{\textbf{FULL}} & \color{brown}{\textbf{FULL}} \\
    \color{brown}{\textbf{FULL}} & WLS & \color{brown}{\textbf{OLS}}
    & FULL & FULL & FULL
    & WLS & \color{brown}{\textbf{OLS}} & OLS \\
    \color{brown}{\textbf{OLS}} & BASE & PBU
    & WLS & BASE & BASE
    & OLS & BASE & BASE \\
    \color{brown}{\textbf{PBU}} & PBU & BASE
    & OLS & PBU & PBU
    & PBU & PBU & PBU \\
    \bottomrule
    \end{tabular}}
    \caption{Outcome of the sequential studentized MCS procedure for the high-dimensional simulation study across $n_b \in \{10, 20, 50\}$, at \(\alpha=0.05\). Bold brown entries indicate the methods retained in the final model confidence set. Eliminated methods are ordered according to the sequential MCS elimination path, with the first eliminated method shown at the bottom.}
    \label{tab:sim_mcs_membership}
\end{table}

\section{Complexity analysis for computation times}
\label{app:comp_time}

\begin{figure}
     \centering
     \includegraphics[width=1\linewidth]{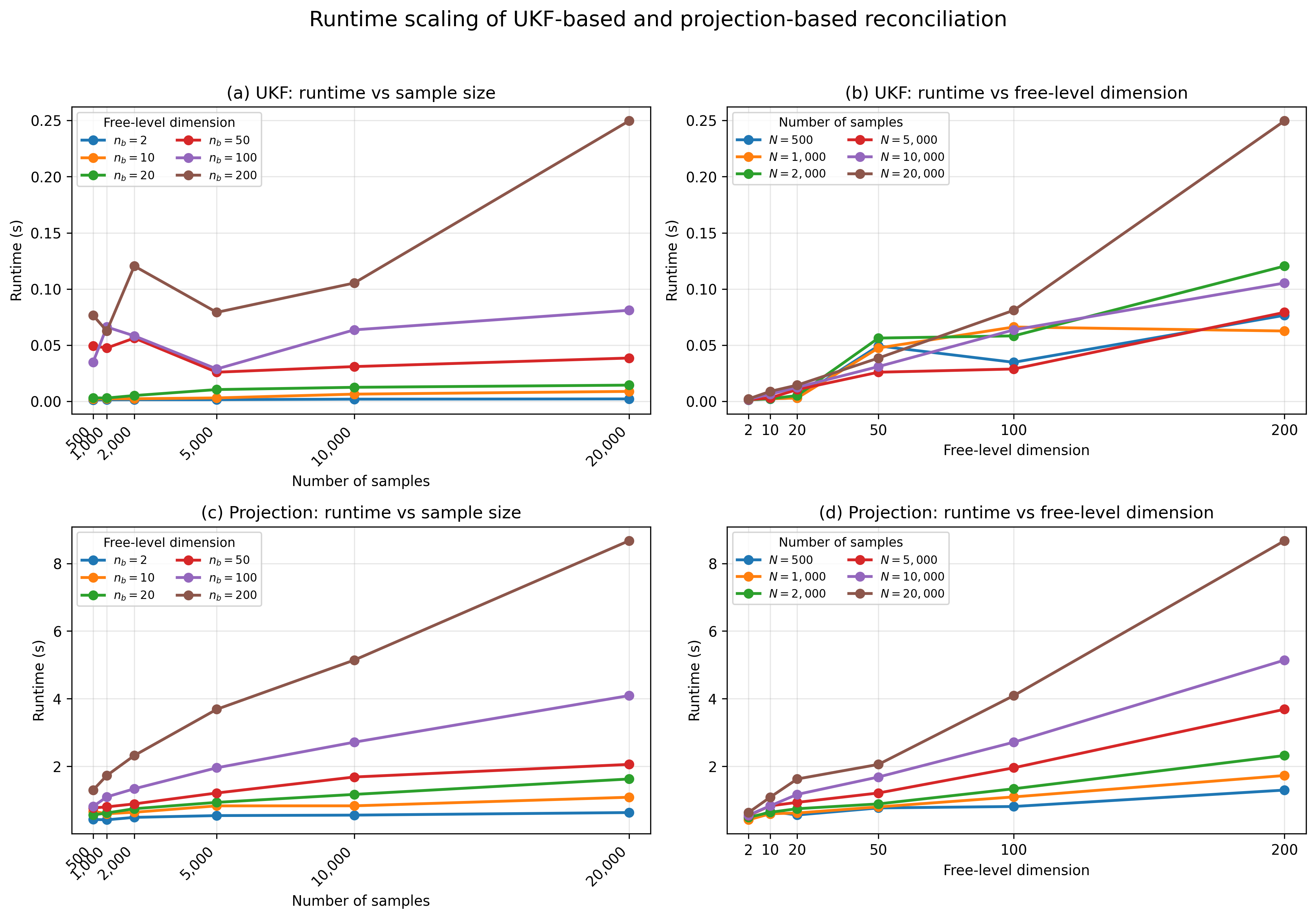}
     \caption{Computation times for the UKF and projection against the dimension of free time series and number of forecasted samples.}
     \label{fig:sim_runtimes}
\end{figure}

\figref{fig:sim_runtimes} reports the empirical runtime behaviour of the projection-based and UKF-based reconciliation methods. For reconciliation via projection, the computation time increases almost linearly with the number of forecast samples. This is expected because each forecast sample must be projected onto the nonlinear constraint manifold. The runtime also increases with the number of free time series, reflecting the greater numerical difficulty of the projection problem in higher-dimensional spaces. These trends are empirical and should not be interpreted as formal asymptotic complexity results, since the actual cost depends on the specific numerical optimization algorithm used for the projection step.

For reconciliation via conditioning, the runtime is much less sensitive to the number of forecast samples. The UKF-based method approximates the reconciled conditional distribution as a Gaussian distribution; its main computational cost is therefore driven by the construction and propagation of sigma points. Since the Unscented Transform uses \(2n_b+1\) sigma points, where \(n_b\) is the number of free variables, the runtime increases with the number of free time series. By contrast, the number of samples mainly affects the final sampling stage, where draws are generated from the reconciled Gaussian distribution and mapped through the FTA transformation. In the experiments, this final stage has a relatively small impact on the total runtime, except at the largest sample size \(N=20000\), where the sampling cost becomes more visible.

\bibliography{refs}

\end{document}